\documentclass[aps,prd,twocolumn,showpacs,superscriptaddress,groupedaddress]{revtex4}  % for review and submission
\usepackage{graphicx}  % needed for figures
\usepackage{dcolumn}   % needed for some tables
\usepackage{bm}        % for math
\usepackage{amssymb}   % for math
\usepackage{slashed}
\begin{document}

% The following information is for internal review, please remove them for submission

% the following line is for submission, including submission to the arXiv!!
%\hspace{5.2in} \mbox{Fermilab-Pub-04/xxx-E}

\title{Beyond fast rate approximations: General analytic solutions to coupled transport equations during cosmic phase transitions}

\author{G. A.  White}
\affiliation{ARC Centre of Excellence for Particle Physics at the Tera-scale \\
School of Physics and Astronomy,
Monash University\\
Vicotria 3800, Australia}
\date{\today}
\pacs{98.80.-k, 11.30.Fs, 05.30.Rt,11.30.Pb}
\email{graham.white@monash.edu}
\begin{abstract}
We propose a general method to analytically solve transport equations during a cosmic phase transition without making approximations based on the assumption that any transport coefficient is large. Using the MSSM as an example we derive the solutions to a set of $3$ transport equations derived under the assumption of supergauge equilibrium and the diffusion approximation. The result is then rederived efficiently using a technique we present involving a parametrized ansatz which turns the process of deriving a solution into an almost elementary problem. We then show how both the derivation and the parametrized ansatz technique can be generalized to solve an arbitrary number of transport equations.  Finally we derive a perturbative series that relaxes the usual approximation that inactivates VEV dependent relaxation and CP violating source terms at the bubble wall and through the symmetric phase.  Our analytical methods are able to reproduce a numerical calculation in the literature.
\end{abstract}
\maketitle
\section{Introduction}
The phase history of our Universe is unknown as it depends on the full content of the scalar potential \cite{cosmicphase}. The many possibilities have inspired a multitude of mechanisms to produce the observed asymmetry between particles and antiparticles during a phase transition \cite{multitude1}--\cite{multitude4}.
The observed Baryogenesis of the universe (BAU)  is estimated to be \cite{wmap}
\begin{equation}
Y_B = \frac{n_B}{s}= \left\{\begin{array}{c} (7.3 \pm 2.5) \times 10^{-11} , \ BBN \\ (9.2 \pm 1.1 )\times 10^{-11} , \ WMAP
\end{array} \right.
\end{equation}
Although the methods derived in this work have general application, we will be working most closely with the electroweak-baryogenesis picture. In the electroweak baryogenesis picture, the BAU is produced during the electroweak phase transition \cite{electroweakbaryogenesis}. To avoid electroweak sphalerons from washing out any produced electroweak baryogenesis one requires this phase transition to be strongly first order \citep{phasetransition}.  The particle dynamics become non-Markovian during the far from equilibrium conditions that occur during a cosmic phase transition rendering equilibrium quantum field theory inadequate as a tool to analyse the behaviour of particle number densities throughout. One instead is required to use the closed time path (CTP) formalism \cite{CTPformalism1}--\cite{CTPformalism6}. The usual treatment involves deriving a set of coupled quantum transport equations (QTE) involving CP conserving relaxation terms and CP violating sources derived from scattering amplitudes evaluated using the CTP formalism \cite{CPVsource}--\cite{relaxation}. There are in general as many QTEs as there are particle species in your model so one may be left with a daunting number of coupled inhomogeneous differential equations to solve. Therefore one usually proceeds either by solving the system numerically or by using a set of approximations to reduce the number of QTEs to a single one which is then solved in two regions with the solutions matched at the phase boundary. The former method is very cumbersome and the latter method involves using a set of fast rate approximations to reduce the set of coupled transport equations down to one (e.g. \cite{relaxation} and the references therein for this treatment in the MSSM). However, recently it has been shown that in the case of the minimal supersymmetric standard model (MSSM) with $R$ parity conservation, the assumption of fast Yukawa and triscalar rates is infrequently justified \cite{triscalar}. Furthermore, the practice of solving the equations in two regions effectively approximates the VEV profile with a step function and ignores any effects of a thick bubble wall, an assumption we refer to in this paper as the ``ultra thin wall approximation''. 
 In this work we show how to analytically  go beyond these approximations by deriving an exact solution to a general set of QTEs. We begin by considering the set of QTEs that govern the MSSM during an electroweak phase transition derived under the assumption of local supergauge equilibrium  and null number densities for weak bosons. This is admittedly not an exact system to begin with but it is useful as an example before we generalizing our method to larger systems of coupled transport equations with multiple CP violating source terms. We also show how to analytically derive an expansion that goes beyond the ultra thin wall approximation.  Moreover, we show that once one knows the form of the solution - which has a general form - one can more quickly derive the solution using a parametrized ansatz whose parameters are determined by direct substitution. The layout of our paper is as follows. In section \ref{CTP} we review the closed time path formalism and its use in deriving a set of coupled QTEs that govern the behaviour of particle densities across a phase transition. In section \ref{MSSM} we use the two Higgs doublet model (or equivalently the MSSM under the assumption of supergauge equilibrium and a single CP violating source term) as an illustrative example of how to derive an analytic solution to a set of QTEs without using a fast rate approximation. In section \ref{ansatz} we show how to re derive this solution quickly using a parametrized ansatz.  We then generalize our methods to an, in principle, arbitrary set of transport equations with multiple CP violating sources in section \ref{generalizations}. Before showing how to go beyond the ultra thin wall approximation in section \ref{ultra}. Finally in section \ref{numerics} we compare our analytic solution to what one might get using the fast rate approximation for one set of parameters as well as numerically calculating the lowest order correction to the ultra thin wall regime before concluding in section \ref{conclusion}.
\section{Closed Time Path Formalism}\label{CTP}

The evolution of quantum states at finite temperature and far from equilibrium is qualitatively different to zero temperature quantum field theory in that the equilibrium relation between in and out states is broken. It is therefore appropriate to use the ``Closed Time Path formalism'' (CTP) to study this evolution. For the CTP formalism our time contour starts $\epsilon $ above the real line from the infinite past to the present before travelling to a point $\epsilon$ below the real line and then the infinite past. Finally, for technical reasons, the contour goes a final journey orthogonal to the real line to the point $t=-\infty - i \beta$. In this formalism one can begin with the Schwinger-Dyson equation for a particular quantum field to derive an equation that relates the divergence of the current to its self energies
\begin{eqnarray}
\partial _\mu J^\mu &=&- \int _{-\infty } ^{X_0} dz^0 \int d^3z  {\rm Tr}  \left[\Sigma ^> (x,z) G^< (z,x) \right. \nonumber \\ &&- \left. G ^> (x,z) \Sigma ^< (z,x) +G ^< (x,z) \Sigma ^> (z,x) \right. \nonumber \\ && \left. - \Sigma ^< (x,z) G ^> (z,x) \right]\ .   \label{fermionsource}
\end{eqnarray}
Here, $G^\lambda $ are the finite temperature fermionic propagators
\begin{eqnarray}
G^>(x,z) &=& \langle  \phi _-(x)  \phi ^\dagger _+(z) \rangle \nonumber \\
G^<(x,z) &=& \langle \phi ^\dagger _- (x) \phi _+(z) \rangle  
\end{eqnarray}
and $\Sigma ^\lambda $ are the self energies.  We can now construct source and relaxation terms using the closed time path Feynman rules. The simplest example, as shown in Fig 1., involves a particle scattering with a space-time varying VEV.   As an example consider the interaction of the top squark in the MSSM with the space time varying VEV. 
\begin{equation}
L \ni y_tA_t v_u \tilde{t} _L \tilde{ t }_R -\mu ^* y_t v _d \tilde{t }_L \tilde{ t } _R 
\end{equation}
As an approximation we can use the mass eigenbasis for the symmetric phase and treat the interactions with Higgs perturbatively.  The self energy of this diagram is given by
\begin{eqnarray}
&& \tilde{\Sigma} _R (x,z) = \nonumber \\ &&-y_t^2 \left[A_t v_u(x)-\mu ^* v_d(x) \right] \left[A_t^* v_u(z)-\mu  v_d(z) \right]\tilde{G} ^0_L(x,z) \nonumber \\
\end{eqnarray}
where $v_u=\sin \beta v$, $v_d=\cos \beta v$ and $v^2=v_u^2 +v_d^2$.
We can use the identity $\left[G^\lambda (x,z)\right]^*=G^\lambda (z,x)$ to write Eq. [\ref{fermionsource}] as a sum of $CP$ conserving relaxation terms and CP violating source terms. We can write the CP conserving relaxation term as
\begin{equation} \sum _\pm \Gamma ^\pm _M (\mu _{L} \pm \mu _{R} ) 
\end{equation}
with
\begin{eqnarray}
\Gamma ^{\pm } _M  &=& -\frac{1}{T} \frac{N_Cy_t^2}{4 \pi ^2} |A_tv_u(x)-\mu ^* v_d(x)|^2 \nonumber \\ && \int \frac{k^2dk}{ \omega _b \omega _{L}}  {\rm Im} \left[( {\cal E}_L{\cal E}_b+k^2) \left(\frac{h_F({\cal E}_b)\mp h_F({\cal E}_L)}{({\cal E}_b +{\cal E}_L)} \right) \right. \nonumber \\  &&\left. +({\cal E}_L^*{\cal E}_b-k^2) \left(\frac{h_F({\cal E}_b) \mp n_F({\cal E}_L^*)}{({\cal E}_b -{\cal E}_L^*)} \right) \right] \ .
\end{eqnarray}
In the above equation the frequency is shifted by a thermal width. That is ${\cal E}_x \equiv \omega _x + i \Gamma _x$.  As usual we can make the approximation that $\Gamma ^- _M >> \Gamma ^+_M$ and ignore the smaller term.  Here ${\cal E} _x$ is given by the frequency plus the thermal width $i \Gamma _x$. The CP violating source term is given by
\begin{eqnarray}
S^{\slashed{CP}} &=&\frac{N_Cy_t^2}{2 \pi ^2} {\rm Im} (\mu A_t)v(x)^2 \dot{\beta }(x) \int \frac{k^2dk}{ \omega _b \omega _{L}}  \nonumber \\ && 
{\rm Im} \left[({\cal E}_L{\cal E}_b+k^2) \left(\frac{n_F({\cal E}_b)+n_F({\cal E}_L)-1}{({\cal E}_b +{\cal E}_L)^2} \right)  \right. \nonumber \\ && \left. +({\cal E}_L^*{\cal E}_b-k^2) \left(\frac{n_F({\cal E}_b)-n_F({\cal E}_L^*)}{({\cal E}_b -{\cal E}_L^*)^2} \right) \right] \ .
\end{eqnarray}

\begin{figure}[htbp]
	\begin{center}
		\includegraphics[width=0.4\textwidth]{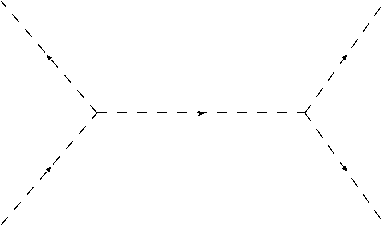}
		\put(-210,125){$v(x)$}
\put(00,125){$v(y)$}
\put(-210,-10){$\tilde{t}_R$}
\put(0,-10){$\tilde{t}_R$}
\put(-105,65){$\tilde{t}_L$}
 %depending on the latex compiler, you can omit the file extension
		\caption{Self energy arising from stop interaction with VEVs}
		\label{vevdiagram}
	\end{center}
\end{figure}
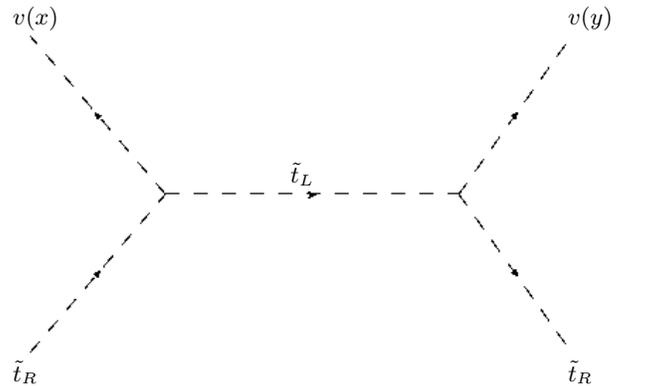
In general, relaxation terms involving chemical potentials $\mu _i$ can be written as a rate times a linear combination of chemical potentials.
Furthermore, the chemical potentials can be related to the number densities through the identity
\begin{equation}
n_i =g_i\int _0 ^\infty  \frac{d^3k}{(2 \pi )^3}\left[ N(\omega _k, \mu_i)-N(\omega _k, -\mu _i)\right] \ ,
\end{equation}
where $N( \omega , \mu )$ is the usual equilibrium distribution for fermions/bosons and $g_i$ is the number of degrees of freedom. Expanding in the chemical potential and ignoring higher order terms we get the relation
\begin{equation}
n_i = \frac{k_i(m_i/T)T^2}{6}T^2
\end{equation}
with
\begin{equation}
k_i(m_i/T)=k_i(0)\frac{c_{F,B}}{\pi ^2}\int ^\infty _{m/T} x dx \frac{e^x}{(e^x\pm 1)^2} \sqrt{x^2-m^2/T^2}
\end{equation}
with $c_{F,B}=6(3)$ for fermions (bosons) and $k_i(0)=1$ for chiral fermions and $2$ for Dirac fermions and complex scalars. Finally the denominator has a $+$ for fermions and a $-$ for bosons. 
\section{The MSSM: an illustrative example}\label{MSSM}
The transport equations for the MSSM under the assumption of supergauge equilibrium as well the assumption that $\mu _W ^+\pm \approx 0$ has been derived many times before (see for instance \cite{relaxation} and the references therein). So we just quote the result here. Under the assumption of supergauge equilibrium we only need to write coupled transport equation for the following number densities
\begin{eqnarray}
Q&=&n_{t_L}+n_{b_L}+n_{\tilde{t}_L}+n_{\tilde{b}_L} \nonumber \\
T&=& n_{t_R} + n_{\tilde{t} _R} \nonumber \\ 
H &=& n_{H_u ^+}+n_{H_u ^0} -n_{H_d ^-}-n_{H_d ^0} \nonumber \\ && n_{\tilde{H}_u ^+}+n_{\tilde{H}_u ^0} -n_{\tilde{H}_d ^-}-n_{\tilde{H}_d ^0} \ .
\end{eqnarray}
The set of coupled differential equations are
\begin{eqnarray}
\partial _ \mu T^\mu &=& \Gamma _M ^+ \left(\frac{T}{k_T}+\frac{Q}{k_Q} \right)-\Gamma _M ^- \left(\frac{T}{k_T}-\frac{Q}{k_Q} \right) \nonumber \\ &&- \Gamma _Y \left( \frac{T}{k_T}-\frac{H}{k_H}-\frac{Q}{k_Q} \right) \nonumber \\ && +\Gamma _{SS} \left(\frac{2Q}{k_Q}-\frac{T}{k_T}+\frac{9(Q+T)}{k_B} \right)+ S^{\slashed{CP}} _{\tilde{t}} \nonumber \\ 
\partial _ \mu Q^\mu &=& - \Gamma _M ^+ \left(\frac{T}{k_T}+\frac{Q}{k_Q} \right)-\Gamma _M ^+ \left(\frac{T}{k_T}-\frac{Q}{k_Q} \right) \nonumber \\ &&+ \Gamma _Y \left( \frac{T}{k_T}-\frac{H}{k_H}-\frac{Q}{k_Q} \right) \nonumber \\ &&-2\Gamma _{SS} \left(\frac{2Q}{k_Q}-\frac{T}{k_T}+\frac{9(Q+T)}{k_B} \right)- S^{\slashed{CP}} _{\tilde{t}} \nonumber \\ 
\partial _ \mu H^\mu &=&-\Gamma _H \frac{H}{k_H}+ \Gamma _Y \left(\frac{T}{k_T}-\frac{Q}{k_Q} -\frac{H}{k_H} \right)+S_{\tilde{H}}^{\slashed{CP}} \ . \nonumber \\
\end{eqnarray}
Our strategy will involve writing the above equation in what we call ``cascading form'' where the first equation is a function of just two number densities, the second a function of three and the third also a function of three number densities as well as the CP violating source. In our example it is easy to write this set of equations in such a form, the only complication being the existence of two space-time dependant source terms. To overcome this notice that both sources have the same space time dependence up to a overall constant of proportionality
\begin{equation}
S_{\tilde{t}} ^{\slashed{CP}} = \frac{1}{2a} S_{\tilde{H}} ^{\slashed{CP}} \ .
\end{equation}
It is therefore straightforward to take a set of linear combinations that are in cascading form
\begin{eqnarray}
&& \partial _\mu (T+Q)^\mu =-\Gamma _{SS} \left(\frac{2Q}{k_Q}-\frac{T}{k_T}+\frac{9(Q+T)}{k_B} \right) \nonumber \\
&& \partial _\mu (T(1+a)+Q(1-a)+H)^\mu = \nonumber \\ && (1-2a) \Gamma _Y \left(\frac{T}{k_T}-\frac{Q}{k_Q} -\frac{H}{k_H} \right) \nonumber \\ &&-(1-3a)\Gamma _{SS} \left(\frac{2Q}{k_Q}-\frac{T}{k_T}+\frac{9(Q+T)}{k_B} \right) -\Gamma _H \frac{H}{k_H} \nonumber \\ && +2a \Gamma _M ^+ \left(\frac{T}{k_T}+\frac{Q}{k_Q} \right)-2a \Gamma _M ^- \left(\frac{T}{k_T}-\frac{Q}{k_Q} \right)\nonumber \\ 
&& \partial _\mu (2T+Q+H)^\mu = \Gamma _M ^+ \left(\frac{T}{k_T}+\frac{Q}{k_Q} \right)-\Gamma _M ^- \left(\frac{T}{k_T}-\frac{Q}{k_Q} \right)\nonumber \\ && -\Gamma _H \frac{H}{k_H} +(1+2a)S^{\slashed{CP}} _{\tilde{t}} \ . \label{MSSMcascade}
\end{eqnarray} 
If we use the standard diffusion approximation \begin{equation} \partial _\mu J^\mu = v_W \dot{n} -D_n \nabla ^2 n \end{equation} plus make an assumption about the geometry of the bubble wall, we can reduce the problem to a one-dimensional one. Shifting to the bubble wall rest frame using the variable $z\equiv |v_wt-x|$, we can write the set of coupled transport equations as a set of differential equations in $z$.
A strategy for a solution becomes immediately apparent when the equations are written in cascade form. The first equation is a differential operator acting on $T$ and $Q$ which can be solved for $T$ (or equivalently $Q$) by treating $Q$ and its derivatives as a inhomogeneous source. Thus we can write the second equation in terms of $H$ and $Q$ which can also be solved in a similar way so that $H$ is a function of $Q$. The third and final equation is then an equation for $Q$ and the CP violating source which can be solved via the usual methods.
\subsection{Deriving the analytic solution}
Let us rewrite Eq. (\ref{MSSMcascade} ) in terms of implicitly defined coefficients $a_{Xj}^i$ where $X \in \{Q,T,H \}$ is the field indice, $j \in \{1,2,3 \}$ is the equation number and $i \in \{0,1,2\}$ is the power of the derivative in $z$
\begin{eqnarray}
a_{Q1}^i \partial ^i Q+a_{T1}^i \partial ^i T&=& 0 \\ \label{cascade1}
a_{Q2}^i \partial ^i Q+a_{T2}^i \partial ^i T+a_{H2}^i \partial ^i H &=& 0 \\ \label{cascade2}
a_{Q3}^i \partial ^i Q+a_{T3}^i \partial ^i T+a_{H3}^i \partial ^i H &=& \Delta (z) \ . \label{cascade3}
\end{eqnarray}
The above equation and throughout this paper we of course use Einstein's summation convention for repeated indices.
The first equation can be used to solve for $T$ in terms of $Q$ if one uses the method of variable coefficients and treats the part of the equation involving $Q$ and its derivatives as an inhomogeneous source for $T$. The result of this is
\begin{equation}
T =\frac{1}{a_{T1}^2 }\sum _\pm \frac{1}{\kappa _\mp -\kappa _\pm} e^{\kappa _{\pm } z} \left[ \int ^z e^{- \kappa _\pm y} \left( a^i_{Q 1} \frac{\partial ^i Q}{\partial y^i}   \right) dy -\beta _i\right] \label{TinQ}
\end{equation}
where the integration constants have to be zero as explained in appendix \ref{note}. We neglect them for the rest of the derivation as they clutter notation and detract from the main derivation.  The above equation has the problem that if we used it to eliminate $T$ in subsequent equations we would be left with a set of equations that are a mixture of an integral and differential equations. We wish to have a pure differential equation at the end. To achieve this we use a series of variable changes. The derivation is somewhat non-trivial so we give extra detail in the appendix. Here we sketch out the main points of the calculation. We begin with the following change of variables
\begin{equation}
h_\pm = \int ^z e^{- \kappa _\pm y} Q dy
\end{equation}
from which we can eliminate the integral and the exponent in equation (\ref{TinQ}) at the cost of having $T$ now defined in terms of two functions. These functions are related via the identity
\begin{equation}
h_{+ } ^\prime = e^{(\kappa _- - \kappa _+)z }h_- ^\prime \ .
\end{equation}
To remove the exponential outside of the former integral we use an additional change of variables
\begin{equation}
j_{\pm} = e^{\kappa _\pm z} h _\pm \ .
\end{equation} 
It is now possible to write a simple expression relating $Q$ to either $j_\pm$
\begin{equation}
Q=j_+^\prime - \kappa _+ j_+=j_-^\prime - \kappa _- j_- \ .
\end{equation}
We are not quite done, we would like to now write everything in terms of a single variable. This can be achieved by either of the following variables
\begin{equation}
k= e^{\kappa _\mp z} \int ^z e ^{- \kappa _\mp y} j_\pm dy
\end{equation}
from which we can relate both $j_\pm $ to $k$ through the equation
\begin{equation}
j_{\pm} = k^\prime - \kappa _\mp k \ .
\end{equation}
We now have a single variable $k$ and it is straight forward to derive expressions relating $T$ and $Q$ to derivatives in $k$
\begin{eqnarray}
T&=& -\frac{1}{a_{T1}^2}a^i _{Q1}\partial ^i k \\
Q &=& \frac{a^i _{T1}}{a^2_{T1}}\partial ^i k \ .
\end{eqnarray}
For the sake of convenience let us rescale $k$ to remove the denominator and write 
\begin{eqnarray}
T&=& -a^i _{Q1}\partial ^i k \\
Q &=& a^i _{T1}\partial ^i k \ .
\end{eqnarray}
It is a trivial check to verify that these solutions indeed solve the first transport equation.
Substituting these equations into Eq. (\ref{cascade1}) we have a differential equation that is now a function of $k$ and $H$ only, which means we can use the same tricks 
\begin{eqnarray}
0 &=& a^i _{Q2} \partial ^i Q + a^i _{T2} \partial ^i T+a^i _{H2} \partial ^i H \\
&=&  \left( a^i _{Q2} a^j _{T1}- a^i _{T2} a^j _{Q1}\right) \partial ^{i+j} k+a^i _{H2} \partial ^i H \ . \nonumber \\
\end{eqnarray}
The solution for $H$ is of course
\begin{eqnarray}
H &=& \frac{1}{a_{H2}^2 }\sum _\pm \frac{e^{\kappa _{\pm } z} }{\kappa _\mp -\kappa _\pm} \int ^z e^{- \kappa _\pm y} \nonumber \\ && \left( a^i _{Q2} a^j _{T1}- a^i _{T2} a^j _{Q1} \frac{\partial ^{i+j} k}{\partial y^i}   \right) dy
\end{eqnarray}
We can make the exact same changes of variables before to solve this in terms of $l$ (which is analogous to $k$ in the solution to the first equation)
\begin{eqnarray}
H&=& -\sum _{n=0}^4 \delta _{i+j -n} (a^i_{Q2}a^j _{T1}-a^i_{T2}a^j_{Q1})\partial ^nl \\
k &=& a^i _{H2} \partial ^i l
\end{eqnarray}
where the Kronecker delta was added to make the structure of the solution more transparent and we include the sum over $n$ to make its limits obvious. Subbing our solution for $k$ into the equations for $Q$ and $T$ gives
\begin{eqnarray}
H &=& -\sum _{n=0}^4 \delta _{i+j-n} (a^i _{Q2} a^j_{T1} - a^i _{T2} a^j_{Q1}) \partial ^n l \nonumber \\
T &=& -\sum _{n=0}^4 \delta _{i+j-n} a^i _{Q1} a^j _{H2}  \partial ^nl \nonumber \\
Q &=& \sum _{n=0}^4 \delta _{i+j-n} a^i _{T1}a^j_{H2}  \partial ^nl \ .
\end{eqnarray}
Equation (\ref{cascade3}) is now defined completely in terms of $l$ and the source
\begin{eqnarray}
\Delta (z) &=& \sum _{n=0}^6 \delta _{i+j+k-n} \left( a^i _{T1} a^j_{H2} a^k _{Q3}-a^i _{Q1} a^j_{H2} a^k _{T3} \right. \nonumber \\ && \left. -a^i _{T1} a^j_{Q2} a^k _{H3}+a^i _{Q1} a^j_{T2} a^k _{H3} \right) \partial ^nl \nonumber \\ 
&=& \sum _{n=0} ^6 \delta _{i+j+k-n} \epsilon ^{abc} a^i_{Ta}a^j _{Hb}a^k_{Qc} \partial ^n l  \nonumber \\  & \equiv &  \sum _{n=0} ^6 a _l ^n \partial ^n l \ .
\end{eqnarray}
The use of the permutation symbol arises from the fact that $a^i_{H1}=0$. The above is a straight forward inhomogeneous differential equation which can be solved using the standard method of variable constants. The solution is
\begin{equation}
l = \sum _{i=1}^6 x_i e^{\alpha _i z} \left( \int ^z e^{-\alpha _i y} \Delta (y) dy - \beta _i  \right) \ 
\end{equation}
in the broken phase and 
\begin{equation}
l = \sum _{i=1}^6 y_i e^{\gamma _i z}  \ 
\end{equation}
in the symmetric phase.
Here the exponents $\alpha _i $  and  $\gamma _i$ are the roots of the equations
\begin{eqnarray}
\sum _{n=0} ^6 a_l ^n \alpha ^n&=&0 \nonumber \\
\sum _{n=0} ^6 a_l ^n \gamma ^n&=&0 \ ,
\end{eqnarray}
for the broken and symmetric phase respectively,
the values $x_i$ can be derived from the equation
\begin{equation}
\vec{x} = M^{-1} \vec{d}  \  .
\end{equation}
The matrix $M$ is given by $M_ij \equiv \alpha ^{j-1} _i $ where $j$ doubles as an exponent and an indices which both go from $1$ to $6$ and $\vec{d}  \equiv [0,\cdots , 1 /a_l^6 ]^{\rm T}$. Finally the integration constants $\beta _i $
and $y_i$ are determined by the boundary conditions as follows. Since we began with a set of three second order differential equations for three densities we require that each of these number densities be continuous at the bubble wall along with their derivative. We also require the number densities to vanish at $\pm \infty $. All exponents therefore have to be positive definite in the symmetric phase. Therefore
\begin{equation}
y_i =0 \ \forall \gamma _i \leq 0 \ .
\end{equation}
To ensure number densities go to zero in the broken phase we have a condition on all positive definite exponents
\begin{equation} 
x_i \beta _i = x_i \int ^\infty _0 dy e^{-\alpha _iy} \Delta (y) \equiv I_i \ \forall \alpha _i \geq 0 \ .
\end{equation}
Finally our continuity conditions need to be met. The continuity conditions are not met by requiring $l$ and its derivatives to be continuous at the phase boundary even though this will give the right number of conditions. Instead we require that $H,T,Q,H^\prime ,T ^\prime $ and $Q^\prime $ are continuous at $z=0$. As an example consider the case when the last three $\gamma _i$ exponents are greater than $0$ as well as the first three $\alpha _i$ exponents. The continuity conditions can be met when the following equation is satisfied \begin{widetext}
\begin{eqnarray}
&& \left( \begin{array}{ccccccccc} x_4 \beta _4 & x_5 \beta _5 & x_6 \beta _6 & x_1 \beta _1 & x_2 \beta _2 & x_3 \beta _3 & y_1 & y_2 & y_3 
\end{array} \right) ^T = \nonumber \\ && \left( \begin{array}{ccccccccc} 1 & 0 & 0 & 0&  0 & 0 & 0 & 0 & 0 \\ 0 & 1 & 0&  0&  0 & 0 & 0 & 0 & 0 
 \\ 0 & 0 & 1&  0&  0 & 0 & 0 & 0 & 0   \\ A_Q(\alpha _4)  & A_Q(\alpha _5)  & A_Q(\alpha _6)   & A_Q(\alpha _1)  & A_Q(\alpha _2)  & A_Q (\alpha _3) &  A_Q(\gamma _1) & A_Q(\gamma _2) &  A_Q(\gamma _3) \\ \alpha _4 A_Q(\alpha _4)  & \alpha _5 A_Q(\alpha _5)  & \alpha _6 A_Q(\alpha _6)  & \alpha _1  A_Q(\alpha _1)  & \alpha _2 A_Q(\alpha _2)  & \alpha _3 A_Q (\alpha _3) & \gamma _1 A_Q(\gamma _1) & \gamma _2 A_Q(\gamma _2) & \gamma _3 A_Q(\gamma _3)  \\ A_T(\alpha _4)  & A_T(\alpha _5)  & A_T(\alpha _6)   & A_T(\alpha _1)  & A_T(\alpha _2)  & A_T (\alpha _3)&  A_T(\gamma _1) &  A_T(\gamma _2) &  A_T(\gamma _3) \\ \alpha _4 A_T(\alpha _4)  & \alpha _5 A_T(\alpha _5)  & \alpha _6 A_T(\alpha _6)  & \alpha _1  A_T(\alpha _1)  & \alpha _2 A_T(\alpha _2)  & \alpha _3 A_T (\alpha _3) & \gamma _1 A_T(\gamma _1) & \gamma _2 A_T(\gamma _2) & \gamma _3 A_T(\gamma _3) \\ A_H(\alpha _4)  & A_H(\alpha _5)  & A_H(\alpha _6)   & A_H(\alpha _1)  & A_H(\alpha _2)  & A_H (\alpha _3) & A_H(\gamma _1) & A_H(\gamma _2) &  A_H(\gamma _3) \\ \alpha _4 A_H(\alpha _4)  & \alpha _5 A_H(\alpha _5)  & \alpha _6 A_H(\alpha _6)  & \alpha _1  A_H(\alpha _1)  & \alpha _2 A_H(\alpha _2)  & \alpha _3 A_H (\alpha _3) & \gamma _1 A_H(\gamma _1) & \gamma _2 A_H(\gamma _2) & \gamma _3 A_H(\gamma _3)
\end{array} \right) ^{-1} \left( \begin{array}{c} I_1 \\ I_2 \\ I_3 \\ 0 \\ 0  \\ 0  \\ 0  \\ 0  \\ 0 \end{array} \right) \nonumber \\ &\equiv & \left( \begin{array}{cc} \bf{1}_{3 \times 3} & 0 \\ \vec{A}_X(\alpha )& \vec{A}_X(\gamma ) \\ \vec{( \alpha  A)}_X(\alpha )& \vec{( \alpha  A)}_X(\gamma ) \end{array} \right) ^{-1}  \left(  \begin{array}{c} I_1 \\ I_2 \\ I_3 \\ 0 \end{array} \right)
\end{eqnarray}
\end{widetext}
It is now straight forward to derive the solutions to $T,H$ and $Q$ 
\begin{eqnarray}
H &=& \sum _{i=1}^6 A_H (\alpha _i) x_i e^{\alpha _i z} \left( \int ^z e^{-\alpha _i y} \Delta (y) dy - \beta _i  \right) \nonumber \\
T &=& \sum _{i=1}^6 A_T (\alpha _i) x_i e^{\alpha _i z} \left( \int ^z e^{-\alpha _i y} \Delta (y) dy - \beta _i  \right) \nonumber \\
Q &=& \sum _{i=1}^6 A_Q (\alpha _i) x_i e^{\alpha _i z} \left( \int ^z e^{-\alpha _i y} \Delta (y) dy - \beta _i  \right) \nonumber \\
\end{eqnarray}
with known functions
\begin{eqnarray}
A_H &=& -\sum _{n=0}^4 \delta _{i+j-n} (a^i _{Q2} a^j_{T1} - a^i _{T2} a^j_{Q1}) \alpha ^n \nonumber \\
A_T &=& -\sum _{n=0}^4 \delta _{i+j-n} a^i _{Q1} a^j _{H2}  \alpha ^n \nonumber \\
A_Q &=& \sum _{n=0}^4 \delta _{i+j-n} a^i _{T1}a^j_{H2} \alpha ^n \ .
\end{eqnarray}
\section{fast rederivation with a paramaterized Ansatz}\label{ansatz}
The above derivation was cumbersome even for the simple example we used and yet the solution was relatively simple. Now that we know the form of the solution, it is far more efficient to start with this form of the solution as a parametrized ansatz and use the differential equations to work out the parameters of the ansatz. This is achieved by the following simple steps.
Let the ansatz for number density $X$ that solves the homogeneous version of our differential equations be $X=A_x (\alpha ) e^{\alpha z}$. To find these functions, $A_x$, as well as $\alpha $ do the following
\begin{itemize}
\item To determine the functions $A_x(\alpha )$ substitute our solutions into the first two equations to get relations between these functions. Doing so gives
\begin{eqnarray}
A_T &=& \frac{-a^i_{Q1}\alpha ^i}{a_{T1}^j \alpha ^j} A_Q \nonumber \\
A_H &=&  \frac{-(a^i _{Q2}\alpha ^i A_Q+a^i _{T2} \alpha ^i A_T)}{a_{H2}^j \alpha ^j}
\end{eqnarray}
This differs from our earlier expression.
However this just amounts to a rescaling of $x_i$ in our final expression. It is a trivial exercise to reverse this rescaling and verify that the expressions $A_X$ are the same as before.
\item Substitute our functions $A_x (\alpha) e^{\alpha _z}$ into the third equation with the CP violating source switched off. The result is a rational polynomial which can be turned into a polynomial by multiplying through with the denominators
\begin{eqnarray}
0&=&   a_{Q3} ^i \partial ^i Q + A_T(\alpha )a_{T3} ^i \partial ^i T + A_H(\alpha )a_{H3} ^i \partial ^i H \nonumber \\ 
0 &\mapsto & \sum _{n=0}^6 \delta _{i+j+k-n} \left( a^i _{T1} a^j_{H2} a^k _{Q3}-a^i _{Q1} a^j_{H2} a^k _{T3} \right. \nonumber \\ && \left. -a^i _{T1} a^j_{Q2} a^k _{H3}+a^i _{Q1} a^j_{T2} a^k _{H3} \right) \partial ^n e^{\alpha z} \nonumber \\ 
& \equiv &\sum _{n=0} ^6 a _l ^n \partial ^n e^{\alpha z}  \nonumber \\ 0&=&a _l ^n \alpha ^n \ .
\end{eqnarray}
Solving this gives you the set of homogeneous solutions. Repeating the above steps in the symmetric phase with ansatz $X=A_x (\gamma ) e^{\gamma z}$ gives you the solutions in that phase.
\item Finally solve the inhomogeneous equation
\begin{eqnarray}
0 &=& \sum _{n=0}^6 \delta _{i+j+k-n} \left( a^i _{T1} a^j_{H2} a^k _{Q3}-a^i _{Q1} a^j_{H2} a^k _{T3} \right. \nonumber \\ && \left. -a^i _{T1} a^j_{Q2} a^k _{H3}+a^i _{Q1} a^j_{T2} a^k _{H3} \right) \partial ^n l(z) \nonumber \\ 
\end{eqnarray}
\end{itemize}  
The inhomogeneous solution is then trivially found to be
\begin{eqnarray}
X &=& \sum _{i=1}^6 A_X (\alpha _i) x_i e^{\alpha _i z} \left( \int ^z e^{-\alpha _i y} \Delta (y) dy - \beta _i  \right) \label{solved}
\end{eqnarray}
in the broken phase and
\begin{equation}
X = \sum _{i=1}^6 A_X (\gamma _i) y_i e^{\gamma _i z}
\end{equation}
in the symmetric phase as before. The values for $x_i$ can be found by substitution Eq. (\ref{solved}) into the inhomogeneous equation and insisting that the coefficients of any derivatives of $\Delta (z)$ are zero and the coefficient of $\Delta (z)$ on the left hand side is $1$. This gives the familiar expression
\begin{equation}
\vec{x} = M^{-1} \vec{d} 
\end{equation} 
with $M_{ij}=\alpha _i ^{j-1}$ and $\vec{d}^T=[0,\cdots,0,1/a_l^6$ as before. The integration constants $y_i$ and $\beta _i$ can be found as above by insisting that $Q$, $T$ and $H$ are well behaved at $\pm \infty $ and all three functions as well as their derivatives are continuous at the bubble wall. To make this section self contained and stand alone we repeat the result here 
\begin{eqnarray}
&& \left( \begin{array}{c} x_4 \beta _4 \\ x_5 \beta _5 \\ x_6 \beta _6 \\ x_1 \beta _1 \\ x_2 \beta _2 \\ x_3 \beta _3 \\ y_1 \\ y_2 \\ y_3 
\end{array} \right) = \left( \begin{array}{cc} \bf{1}_{3 \times 3} & 0 \\ \vec{A}_X(\alpha )& \vec{A}_X(\gamma ) \\ \vec{( \alpha  A)}_X(\alpha )& \vec{( \alpha  A)}_X(\gamma ) \end{array} \right) ^{-1}  \left(  \begin{array}{c} I_1 \\ I_2 \\ I_3 \\ 0 \end{array} \right) \nonumber \\
\end{eqnarray}
with $\vec{A}_X(\alpha )=(A_X (\alpha _4),A_X (\alpha _5), A_X (\alpha _6), A_X (\alpha _1),$ $A_X (\alpha _2),A_X (\alpha _3))^T$ and $\vec{A}_X(\gamma ) = [A_X(\gamma _1 ),A_X(\gamma _2) , A_X(\gamma _3) ] ^{\rm T}$. Also $\vec{(\alpha A)}_X(\alpha ) = (\alpha _4 A_X (\alpha _4), \cdots )^T$ and $X \in \{Q,T,H\}$.
Remarkably, the above method of solving the set of differential equations analytically is at least as fast as using an approximation that assumes the strong sphaleron and Yukawa rates are fast in order to reduce our set of transport equations to a single differential equation which one then solves.
\section{generalizations}\label{generalizations}
In this section we discuss how far generalizations of the above procedure can go. The generalization of our procedure is trivial to any case where you have only one CP violating source and you can manipulate the transport equations into the cascading form
\begin{eqnarray}
D_1[X_1,X_2] &=& 0 \nonumber \\
D_2[X_1,X_2,X_3 ] &=& 0 \\
\vdots && \nonumber \\
D_{N-1}[X_1,X_2,X_3 \cdots X_N ] &=& 0 \nonumber \\
D_{N-1}[X_1,X_2,X_3 \cdots X_N ] &=& \Delta (z) \ .
\end{eqnarray}
One can just use the parametrized ansatz approach defined in the previous section to very quickly derive the solution. Much more daunting are the cases where one either has multiple CP violating sources or one cannot manipulate the transport equations into the above form. Fortunately in this section we will show that in principle an analytical solution can always be found even in the presence of such complications and the parametrized ansatz approach generally reduces the difficulty of problems greatly.
\subsection{Multiple CP violating sources}
In the MSSM there are several $CP$ violating sources.  Handling the extra CP violating sources however was done by simply noting that they are all proportional to $v(x)^2 \dot{\beta }(x)$. This situation becomes far more non trivial in the case where there are numerous CP violating source terms proportional to different functions of $z$. For example the NMSSM has source terms that are a function of the singlet's VEV profile, $v_S(z)$ as well as the Higgs profiles $v_u(z)$ and $v_d(z)$. 
Consider the set of transport equations from the previous section with an extra $CP$ violating term
\begin{eqnarray}
a_{Q1}^i \partial ^i Q+a_{T1}^i \partial ^i T&=& 0 \\
a_{Q2}^i \partial ^i Q+a_{T2}^i \partial ^i T+a_{H2}^i \partial ^i H &=& \Delta _1 (z) \\
a_{Q3}^i \partial ^i Q+a_{T3}^i \partial ^i T+a_{H3}^i \partial ^i H &=& \Delta _2(z) \ . \label{2CPVs}
\end{eqnarray}
If we solve the first transport equation as before the second equation becomes
\begin{eqnarray}
\Delta _1 (z)&=&(a_{Q2}^i a_{T1}^j -a_{Q1}^i a_{T2}^j) \partial ^{i+j}l + a_{H2}^i\partial ^i H \nonumber \\ 
&=&a_{l2}^i \partial ^il + a_{H2}^i\partial ^i H \ .
\end{eqnarray}
The solution to this equation is of course
\begin{eqnarray}
H&=& -a_{l2} ^i \partial ^i m+\frac{1}{a^2_{H2} } \sum _{\pm } \frac{1}{\kappa _\mp - \kappa _ \pm } e^{\kappa _\pm z} \nonumber \\ && \left(\int ^z e^{-\kappa _\pm y} \Delta _1(y) dy - \beta _\pm ^1 \right) \ ,
\end{eqnarray}
with $\beta _\pm =0$ for reasons given in the appendix.
This solution feeds into the third equation which obtains a more complicated inhomogeneity as a result
\begin{equation} 
a_{m3}^i \partial ^i m+f(\Delta _1) = \Delta _2 \ ,
\end{equation}
with 
\begin{eqnarray}
&& f(\Delta _1 ) = \nonumber \\ 
&&  \sum _ \pm \frac{(a_H^2)^{-1}}{\kappa _\mp - \kappa _\pm } \left[  a^i _{H3}  \kappa _\pm ^i e^{\kappa _\pm z} \left( \int ^z e^{-\kappa _\pm y} \Delta _1(y) dy - \beta _\pm ^1 \right) \right. \nonumber \\ && \left.  +a^{i+1} _{H3} \kappa _\pm ^i\Delta _1(z) +a_{H3}^2 \Delta _1 ^\prime (z) \right] \ . \nonumber \\ 
\end{eqnarray}
The solution now has a very similar structure to before 
\begin{equation}
m = \sum _i x_i e^{\alpha _i z}  \left( \int ^z e^{-\alpha _i y}\left[ \Delta _2(y) - f(\Delta _1) \right] dy - \beta _i \right)
\end{equation}
with $x_i$, $\alpha _i$, $A_X(\alpha _i)$ and $\beta _i$  determined as before. 
So in principle one can use the parametrized ansatz approach to simplify the problem here as well so long as one is careful in deriving what to replace $\Delta (z)$ with.
\subsection{Other equation structures}
The equation structures that we encountered in the MSSM were not the only possible structures that could arise. In this brief subsection we would like to briefly consider other possible equation structures. Consider an equation where the linear terms are missing
\begin{equation}
a_{1j}^1\frac{\partial X_1}{\partial z}+a_{1j}^1\frac{\partial ^2X_1}{\partial z^2}+a_{2j}^1\frac{\partial X_2}{\partial z}+a_{2j}^1\frac{\partial ^2X_2}{\partial z^2}=0 \ .
\end{equation}
The procedure is subtly different to before. The first step is the same as before where we solve  for $X_1$ by treating the derivatives of $X_2$ as inhomogeneous source 
\begin{equation}
X_1 =- \frac{e^{-\frac{a^1_{j1}}{a^2_{j1}}z}}{a_{1j}^2} \int ^z e^{\frac{a^1_{1j}}{a^2_{1j}}y}\left(a_{2j}^1\frac{\partial X_2}{\partial z}+a_{2j}^1\frac{\partial ^2X_2}{\partial z^2} \right)
\end{equation}
Just as before we aim to remove the exponentials and the integral using a series of changes of variables. This time though our variables changes are
\begin{eqnarray}
h(z) &=& \int ^z e^{\frac{a^1_{1j}}{a^2_{1j}} y}\frac{d X_2}{dy}dy \\
j(z) &=& \int ^z e^{-\frac{a^1_{1j}}{a^2_{1j}} y}h(y)
\end{eqnarray}
which leaves us with a very similar form for the solution (after a suitable rescaling)
\begin{eqnarray}
X_1 &=& a_{2j}^1j ^\prime + a_{2j}^1j ^{\prime \prime } \\
X_2 &=& - a_{1j}^1j ^\prime - a_{1j}^1j ^{\prime \prime } \ .
\end{eqnarray}
Another important extension is to consider the case where we cannot reduce the first equation to a differential equation involving only two number densities. In the simplest case our first equation is a differential operator acting on three number densities
\begin{eqnarray}
a_{1j}^i \partial ^i X_1+a_{2j}^i \partial ^i X_2+a_{3j}^i \partial ^i X_3&=&0 
\end{eqnarray}
in which case we can reduce this to an equation involving only two number density by defining auxillary functions $f(z)$ and $g(z)$ as follows
\begin{eqnarray}
X_1(z) &=& g_{12} X_2(z) +g_{13}(z) X_3+f(z) \\
X_2(z) &=& g_{23} X_3(z) +g(z) 
\end{eqnarray}
if $a_{kj}^0\neq 0$ and 
\begin{eqnarray}
X_1(z) &=& g_{13}(z) X_3+f(z) \\
X_2(z) &=& g_{23} X_3(z) +g(z) 
\end{eqnarray}
otherwise. One can then choose $g_{ij}$ such that they cancel the terms $a_{3j}^i \partial ^i X_3$ in the above transport equation. We therefore have two techniques to convert a set of transport equations to cascading form - taking linear combinations of equations and using the method immediately described above.
\subsection{Arbitrary number of equations}
Consider the case where we have $N$ transport equations and a single CP violating source. In general there are $\frac{3}{2}(N-1)(N-2)$ conditions to satisfy to reorganize the equations into a cascading form. The most general extension of the use of auxillary functions is to define the following
\begin{eqnarray}
X_1 &=& \sum _{j=2}^N g_{1j}X_j+f_1 \nonumber \\
\vdots &=& \vdots \nonumber \\
X_i &=& \sum _{j=i+1} ^N g_{ij} X_j +f_i \nonumber \\
\vdots &=& \vdots \nonumber \\
X_{N-1} &=& g_{N-1,N} X_N+f_{N-1} \ .
\end{eqnarray}
This gives us a total of $\frac{1}{2}N(N-1)$ parameters to use. We can also take linear combinations of the $N-1$ transport equations that do not contain the CP violating source to meet another $\frac{1}{2} N(N-1) $ conditions. Unfortunately this brings us short of the required $\frac{3}{2}(N-1)(N-2)$ conditions. In reality this analysis is probably unrealistically pessimistic for two reasons. First, while the initial set of transport equations indeed can have all $N$ fields appearing in the relaxation terms of all equations, there are only two derivative terms initially in each transport equation. Thus it may be possible to make careful manipulations that evade the arguments given above. Secondly, the equations tend to have a lot more structure than our naive analysis let on. For example a number of equations might have the same relaxation term on the right hand side so the combination of two transport equations with this term might remove many coefficients, $a_{ij}^0$, for the price of one. Nevertheless it is usefull to consider an unrealistically pessimistic scenario where all conditions need to be met. The lowest number of densities that are not guaranteed a path that converts our QTEs into a cascading form in this overly pessimistic scenario is $5$. We then have the transport equations
\begin{eqnarray}
a_{ji}^k \partial ^k X_j &=& 0 \ \forall \ i\leq 4 \\
a_{j5}^k \partial ^k X_j &=& \Delta 
\end{eqnarray}
with $a_{ij}^k \neq 0 \ \forall i,j $ and $k$. We will now sketch out how a solution can be derived. Let us define
\begin{eqnarray}
X_3(z) &=& g_{34} X_4(z) +g_{35}(z) X_5+f_{1}^3(z) \\
X_4(z) &=& g_{45} X_5(z) +f_1^4(z) 
\end{eqnarray}
in such a way as to remove $X_5$ from the first equation. Throughout this argument the subscript on $f$ will denote the number of times a variable has been changed and the superscript will denote the original field that was replaced. Repeating this process to remove $f_1^4$ and then the resulting $f^3_2$ allows us to write the first equation in terms of two linear combinations of number densities which can be denoted $f_3 ^1$ and $f_3^2$. We can then solve this equation to write both in terms of the variable $k$ following the same steps as out lined in previous sections. The second through to fifth equations now involve four variables only. One can again use the same trick to eliminate one number density from the second equation leaving us with something in the form -and we are only interested in a general form as we will use a parametrized ansatz approach shortly -
\begin{equation}
a_{2l}^i \partial ^i k +a_{2f(34)}^i \partial ^i f_4 ^3+a_{2f(44)}^i \partial ^i f_4 ^4 = 0 \ .
\end{equation}
One can then define the following changes in variables
\begin{eqnarray}
f_4 ^3 &=& g_{34l} ^0 k+ g_{34l} ^1 k\prime+g_{34l} ^2 k^{\prime \prime } +f_5^3 \\
f_4^4 &=& g_{44l} ^0 k+ g_{44l} ^1 k\prime+g_{44l} ^2 k^{\prime \prime } +f_5^4
\end{eqnarray}
to cancel the second equations dependants on $l$ (actually this gives us a couple too many parameters so we can set these to 1). This equation then can also be solved with the usual techniques to write $f_5^3$ and $f_5 ^4$ in terms of a new variable $l$. The remaining three equations are a function of $k$, $l$ and $f_5^3$ only. We can use a similar trick to remove the dependency on $l$ in the third equation with the variables 
\begin{eqnarray}
f_5^3 &=&  g_{53l}^0 l+g_{53l}^1 l^\prime +g_{53k}^0 l^{\prime \prime }+f_5^4 \\
k &=& g_{kl} l +f_k ^1 \ .
\end{eqnarray}
The remaining equations are now in cascading form and can be solved using the methods prescribed above. The derivation of the solution to this overly pessimistic case sketched above is very cumbersome. However, even in this case the solution is identical to what one would get if one used a parametrized ansatz from the start. Using the parametrized ansatz method we can write down the solutions immediately without setting pen to paper
\begin{eqnarray}
X_j &=& \sum _{i=0}^{10} x_i A_j(\alpha _i ) e^{\alpha _i z} \left( \int ^ze^{-\alpha _i y}\Delta (y) -\beta _i\right)
\end{eqnarray}
with $\alpha _i$ being the $10$ roots to the equation
\begin{equation}
\sum _{j=1}^5 A_j (\alpha ) a_{j5}^k \alpha _i ^k
\end{equation}
with $A_1(\alpha) =1$ and
\begin{eqnarray}
A_2 &=& \frac{-1}{a^i_{21}\alpha ^i} \sum _{j \neq 2} A_j(\alpha ) a_{j1}^i \alpha ^i  \nonumber \\
A_3 &=& \frac{-1}{a^i_{32}\alpha ^i} \sum _{j \neq 3} A_j(\alpha ) a_{j2}^i \alpha ^i \nonumber \\
A_4 &=& \frac{-1}{a^i_{43}\alpha ^i} \sum _{j \neq 4} A_j(\alpha ) a_{j3}^i \alpha ^i \nonumber \\
A_5 &=& \frac{-1}{a^i_{54}\alpha ^i} \sum _{j \neq 5} A_j(\alpha ) a_{j4}^i \alpha ^i \nonumber \\
\end{eqnarray}
which can be rewritten as a matrix equation
\begin{equation}
\left( \begin{array}{c} A_2 \\ A_3 \\ A_4 \\ A_5
\end{array} \right) = -\left( \begin{array}{cccc} a_{21} ^i \alpha ^i
& a_{31} ^i \alpha ^i & a_{41} ^i \alpha ^i & a_{51} ^i \alpha ^i \\ a_{22} ^i \alpha ^i
& a_{32} ^i \alpha ^i & a_{42} ^i \alpha ^i & a_{52} ^i \alpha ^i \\ a_{23} ^i \alpha ^i
& a_{33} ^i \alpha ^i & a_{43} ^i \alpha ^i & a_{53} ^i \alpha ^i \\ a_{24} ^i \alpha ^i
& a_{34} ^i \alpha ^i & a_{44} ^i \alpha ^i & a_{54} ^i \alpha ^i \end{array} \right) ^{-1} \left( \begin{array}{c} a_{11} ^i \alpha ^i \\ a_{12} ^i \alpha ^i \\ a_{13} ^i \alpha ^i \\
a_{14} ^i \alpha ^i 
\end{array} \right) \ .
\end{equation}
Inverting these equations and rescaling by multiplying through by the denominators gives the very simple structure
\begin{eqnarray}
A_1(\alpha _i) &=& \sum _{n=0}^8 \epsilon ^{bcde} a^i _{2b}a^j _{3c}a^k _{4d}a^l _{5e} \delta _{i+j+k+l-n} \alpha _i ^n  \nonumber \\
A_2(\alpha _i) &=&- \sum _{n=0}^8 \epsilon ^{bcde} a^i _{1b}a^j _{3c}a^k _{4d}a^l _{5e} \delta _{i+j+k+l-n} \alpha _i ^n \nonumber \\
A_3(\alpha _i) &=& \sum _{n=0}^8 \epsilon ^{bcde} a^i _{1b}a^j _{2c}a^k _{4d}a^l _{5e} \delta _{i+j+k+l-n} \alpha _i ^n \nonumber \\
A_4(\alpha _i) &=&- \sum _{n=0}^8 \epsilon ^{bcde} a^i _{1b}a^j _{2c}a^k _{3d}a^l _{5e} \delta _{i+j+k+l-n} \alpha _i ^n  \nonumber \\
A_5(\alpha _i) &=& \sum _{n=0}^8 \epsilon ^{bcde} a^i _{1b}a^j _{2c}a^k _{3d}a^l _{4e} \delta _{i+j+k+l-n} \alpha _i ^n \ . 
\end{eqnarray}
The analogous functions in the MSSM also have this form although using a permutation symbol in that case is probably overkill since there would only be two terms contracted with it and one of those terms is zero two out of three times. Nevertheless this seems to be the standard form. The coefficients $x_i$ are determined as before by the equation 
\begin{equation}
\vec{x} = \left[ \alpha _i ^{j-1} \right]^{-1} \vec{d}  
\end{equation}
with $\vec{d} = [0,\cdots , 0, 1/(a_{5m}^{10})]^{\rm T}$ with $a_{5m} ^{10}$ being the coefficient of $\alpha ^{10}$ in the fifth equation once $A_j(\alpha _i)$ has been substituted in. and the boundary conditions are determined by insisting each field is well behaved at $\pm \infty $ and $X_j$ as well as $X_j^\prime$ are all continuous at the bubble wall which means inverting the equations
\begin{eqnarray}
y_i &=& \ \forall \gamma _i \leq 0 \nonumber \\
\beta _i &=& \int _0 ^\infty dy e^{-\alpha _i y} \Delta(y) \ \forall \alpha _i \geq 0 \nonumber \\
0 &=& \sum _{i,j} A_j(\alpha _i)\beta _k +\sum _{n} y_n  \nonumber \\
0 &=& \sum _{i,j} \alpha _i A_j(\alpha _i)\beta _k +\sum _{n} \gamma _n y_n  
\end{eqnarray}
That such a cumbersome problem was reduced to an elementary one shows the power of the parametrized ansatz method. If we had multiple CP violating source terms we would then simply replace $\Delta (z)$ fir $f[\Delta _1 (z),\Delta _2(z) \cdots]$ and solve as before.
\section{Beyond an ultra thin wall approximations}\label{ultra}
Consider the earlier example given in section \ref{MSSM}. There are VEV dependant relaxation and source terms in the third transport equation that are assumed to switch off in the unbroken phase. For a thick walled VEV profile this approximation might become a poor one. We would like to derive an analytical method to go beyond step function VEVs. The second transport equation (\ref{cascade2}) has VEV dependant relaxation terms proportional to $a$. For now let us simplify the situation by assuming that $a$ is very small so we only need to seek corrections to the thin wall approximation in Eq. (\ref{cascade3}). We will return to the more general case later. To do so we define a series of error functions 
\begin{eqnarray}
\Delta (z) &=& \Theta (z) \Delta (z) +(1-\Theta (z) ) \Delta (z) \equiv \Delta _0 (z) +\epsilon (z) \nonumber  \\
a^0 _{l3} (z) &\equiv &  a^0 _{l3} (z) \Theta (z) +a^0 _{l3}(z)\Theta (-z) \nonumber \\ &=& a^0 _{l3}(z_{\rm max}) -[a^0 _{l3} (z_{\rm max})- a^0 _{l3} (z)] \nonumber \\ && + a_{l3}^0 (z) \Theta (-z) \nonumber \\  &=&a^0 _{l3} +\delta a^0 _{l3}(z) \nonumber \\
l(z) &=& l_0 (z) +\delta _1 l(z) + \delta _2 l(z) +\cdots 
\end{eqnarray}
Here $l_0(z)$ solves the original transport equations in the ultra thin wall regime and $\delta _il$ is a correction of order $i$. We will take advantage of the fact that these error functions are finite for all $z$. This allows the possibility of a perturbative expansion. Our third transport equation can be written in terms of the error functions
\begin{eqnarray}
&& a^i _{l3} \partial ^i l_0 + a_{l3} ^i \partial i (\delta _1 l +\delta _2 l + \cdots ) \nonumber \\ && + \delta a^0 _{l3} (z) (l_0 +\delta _1 l +\delta _2 l +\cdots )= \Delta (z)+ \epsilon (z) \ . \nonumber \\    
\end{eqnarray}
 The corrections to $m$ can be found order by order
\begin{eqnarray}
\delta _1 l &=& \sum _{i=0} ^6 e^{\alpha _i z} x_i  \left( \int ^z e^{-\alpha _i y}\left[\epsilon -l_0\delta a^0 _{l3} (z) \right] -\delta _1 \beta _i  \right) \nonumber  \\
\delta _2 l &=& \sum _{i=0} ^6 e^{\alpha _i z} x_i  \left( \int ^z e^{-\alpha _i y}\left[ -\delta _1l \delta a^0 _{l3} (z) \right] -\delta _2 \beta _i  \right) \nonumber  \\
\end{eqnarray}
etc. 
Let us now turn our attention to the slightly more complicated case where we no longer assume $a$ is small. The equation structure we will be left with is very similar to the case when there are multiple CP violating source terms that are not proportional to each other that we considered in section \ref{generalizations}. Let us define a series of error functions as before. The corrections to the densities are
\begin{eqnarray}
H &=& H_0 + \delta _1 H + \delta _2 H +\cdots \nonumber \\
T &=& T_0 + \delta _1 T + \delta _2 T +\cdots \nonumber \\
Q &=& Q_0 + \delta _1 Q + \delta _2 Q +\cdots \ ,
\end{eqnarray}
where $H_0$, $Q_0$ and $T_0$ solve the QTEs in the ultra thin wall approximation. We can then look at terms that are first order only to set up a perturbative expansion. Using the fact that $Q_0,H_0$ and $T_0$ solve the original set of QTEs we can write QTEs purely in terms of their error functions
\begin{eqnarray}
a^i _{T1} \partial ^i \delta _1 T+a ^i _{Q1} \partial ^i \delta _1 Q &=& 0 \nonumber \\
\delta a^0 _{H2} (z) H_0 +\delta a^0 _{Q2} (z) Q_0 +\delta a^0 _{T2} (z) T_0 +&& \nonumber \\ a^i _{H2} \partial ^i\delta _1 H +a^i _{Q2} \partial ^i\delta _1 Q+a^i _{T2} \partial ^i\delta _1 T &=& 0 \nonumber \\ \delta a^0 _{H3} (z) H_0 +\delta a^0 _{Q3} (z) Q_0 +\delta a^0 _{T3} (z) T_0 +&& \nonumber \\ a^i _{H3} \partial ^i\delta _1 H +a^i _{Q3} \partial ^i\delta _1 Q+a^i _{T3} \partial ^i\delta _1 T &=& \epsilon (z) \ . \nonumber \\
\end{eqnarray}
This can be rewritten in the exact same form as the system we solved in section \ref{generalizations} with two CP violating sources (i.e. Eqs. (\ref{2CPVs})) 
\begin{eqnarray}
a^i _{T1} \partial ^i \delta _1 T+a ^i _{Q1} \partial ^i \delta _1 Q &=& 0 \nonumber \\
 a^i _{H2} \partial ^i\delta _1 H +a^i _{Q2} \partial ^i\delta _1 Q+a^i _{T2} \partial ^i\delta _1 T &=& \Delta _1(z) \nonumber \\ a^i _{H3} \partial ^i\delta _1 H +a^i _{Q3} \partial ^i\delta _1 Q+a^i _{T3} \partial ^i\delta _1 T &=& \Delta _2 (z) \ . \nonumber \\
\end{eqnarray}
\section{a numerical comparison}\label{numerics}
In this section we look at how large an error in the baryogenesis is caused by taking the fast rate approximation and the ultrathin wall approximation. We also look to reproduce a numerical result in the literature as a check of our method. Corrections of $O(1/\Gamma _Y)$ were discussed in \cite{triscalar} which found that the corrections to the baryon asymmetry could be of $O(1)$ in the case where $\Gamma _Y$ and $\Gamma _M ^-$ were similar size in the broken phase. Since a numerical analysis on the effects of equilibrating top Yukawa and stop triscalar terms have already been studied in detail we mostly look at fast rate corrections as a check. We are more interested in determining whether corrections that go beyond the ultra thin wall approximation are large. 
In Table \ref{parameters} we give the set of parameters we use for our numerical analysis.
\begin{center}
\begin{table}[!h]
\begin{tabular}{|c|c||c|c||c|c||}
\hline
$D_T$ & $6/T$ & $D_Q$ & $6/100$ & $D_H$ & $110/T$  \\
\hline
 $a$ & $0.05$ & $\Gamma _M ^- $(x) & $0.5v(x)^2/T^2$ & $\Gamma _H (x) $ & $2a \Gamma _M ^-$ \\
\hline 
$\Gamma _Y$ & $1$ & $ v_w $ & $0.05$ & $L_w $ & $100/T$  \\ \hline $v=\sqrt{v_u^2+v_d^2}$ & $100$ & $T$ & $100$ & $\Delta \beta $ & $0.015$   \\ \hline 
\end{tabular}\caption{table of base set of parameters used for our numerical study. The diffusion constants are taken from reference \cite{diffusion}}\label{parameters}
\end{table}
\end{center}
We also need to define an appropriate space-time dependant VEV profile
\begin{eqnarray}
v(x)&=&0.5v (1-\tanh [\frac{3x}{L_w}]) \nonumber \\
\beta (x) &=& 1-0.5 \Delta \beta (1+\tanh [\frac{3x}{L_w}] ) \nonumber \\
\Delta (x) &=&0.025 \beta ^\prime (x) v(x)^2 \ .
\end{eqnarray}
We can then calculate numerical solutions to the densities $Q,T$ and $H$ which allows us to calculate $n_L=Q+Q_{1L}+Q_{2L}=5Q+4T$. This density, $n_L(z)$, acts as a seed for the baryon density which satisfies the equation \cite{weaksphaleronseed1},\cite{weaksphaleronseed2}
\begin{equation}
D_q \rho ^{\prime \prime }-v_w \rho _B ^\prime-{\cal R}(z) \rho _B =\Gamma _{\rm ws} (z) \frac{n_F}{2} n_L(z)
\end{equation}
with $n_F$ the number of fermion families and the relaxation term is given by
\begin{equation}
R(z) = \Gamma _{\rm ws} (z) \left[ \frac{9}{4} \left(1+\frac{n_{\rm sq}}{6} \right)^{-1}+\frac{3}{2} \right]
\end{equation}
and $n_{sq}$ is the number of squark flavours. Not much is lost treating the weak sphaleron rate profile as a step function - that is compared to treating the VEV profile as a step function when calculating source and relaxation terms - although in principle one could use the same techniques as in section \ref{ultra} to derive corrections to this equation as well.  Our sphaleron rates then have the form \cite{weaksphaleron1}--\cite{weaksphaleron1}
\begin{equation}
\Gamma _{\rm ws} =120 T \alpha _{w} ^5 \Theta (-z) 
\end{equation}
and $\Gamma _{ss} = (128/3) T \alpha _S ^4 $ \cite{strongsphaleron} respectively. The baryon asymmetry is then
\begin{equation}
\rho _B = -\frac{n_F \Gamma _{\rm ws} }{2 v_W} \int _{-\infty } ^0 n_L(x) e^{x R/v_W} dx \ .
\end{equation} 
\begin{figure}[htbp]
	\begin{center}
		\includegraphics[width=0.23\textwidth]{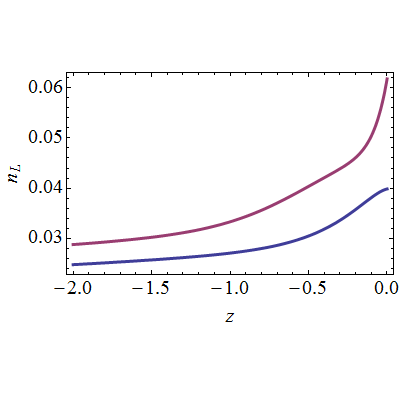} 
		\includegraphics[width=0.23\textwidth]{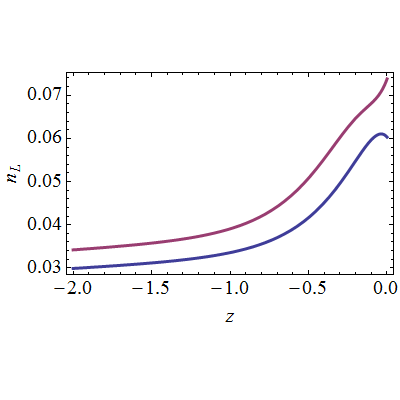} 	\\	
\includegraphics[width=0.23\textwidth]{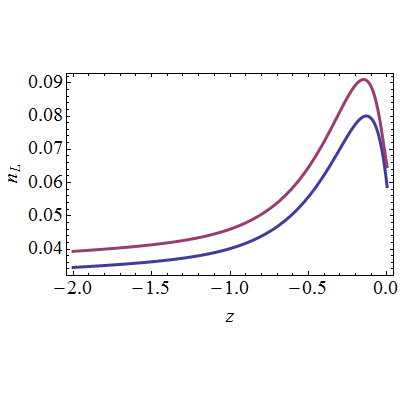}
\includegraphics[width=0.23\textwidth]{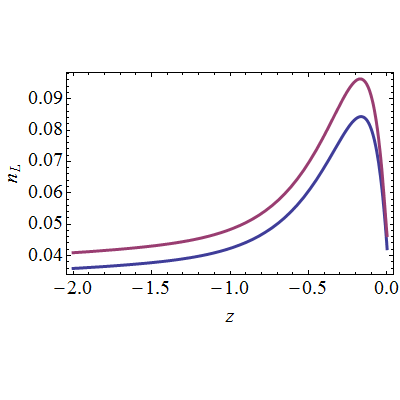}			
		%depending on the latex compiler, you can omit the file extension
		\caption{The density $n_L$ in the symmetric phase near the bubble wall as a function of $z$. The exact solution with first order corrections to the ultra thin wall approximation are given by the magenta line and the solution without such corrections is given in blue.  The parameters used to calculate $n_L(Z)$ are $L_W = 250/T$ (top left), $L_W=100/T$ (top right), $L_W=25/T$ (bottom left) and $L_W=10/T$ (bottom right)}
		\label{corrections}
	\end{center}
\end{figure}
As a point of comparison, we would like to compare our solutions with the analytic solution derived under the fast rate approximation in the ultra thin wall regime \cite{relaxation}. In this case the density $n_L$ \cite{relaxation} with the first $1/\Gamma _{SS}$ correction \cite{gammasscorrections} is given by
\begin{equation}
n_L(z)=-H\left[r_1+r_2\frac{v_w^2}{\Gamma _{SS} \overline{D}} \left(1-\frac{D_q}{\overline{D}} \frac{•}{•} \right) \right]
\end{equation}
with 
\begin{eqnarray}
r_1 &=& \frac{9 k_Qk_T-5k_Qk_B-8k_Tk_B}{k_H(9k_Q+9k_T+k_B)} \nonumber \\
r_2 &=& \frac{k_B^2(5k_Q+4k_T)(k_Q+2k_T))}{k_H(9k_Q+9k_T+k_B)} \nonumber \\
\overline{D} &=& \frac{(9k_Qk_T+k_Bk_Q+4k_Tk_B)D_q}{9k_Qk_T+k_Bk_Q+4k_Tk_B+k_H(9k_Q+9k_T+k_B)} \nonumber \\ && +\frac{k_H(9k_T+9k_Q+k_B)D_h}{9k_Qk_T+k_Bk_Q+4k_Tk_B+k_H(9k_Q+9k_T+k_B)} \nonumber \\
H&=& \frac{e^{v_wz/\overline{D}}}{\overline{D} \kappa _+} \int _0 ^\infty \overline{S}(y)e^{\kappa _+ y}dy \nonumber \\
\kappa _+ &=& \frac{v_w+\sqrt{v_w^2+4\overline{\Gamma } \overline{D}}}{2\overline{D}} \nonumber \\
\overline{\Gamma} &=& \frac{(9k_Q+9k_T+k_B)(\Gamma ^- _M)+\Gamma _h}{9k_Qk_T+k_Bk_Q+4k_Tk_B+k_H(9k_Q+9k_T+k_B)} \nonumber \\
\overline{S} &=&\frac{k_h(9k_Q+9k_T+k_B) (S_{\tilde t}^{\slashed{CP}}+S_{\tilde H}^{\slashed{CP}})}{9k_Qk_T+k_Bk_Q+4k_Tk_B+k_H(9k_Q+9k_T+k_B)}. \nonumber \\
\end{eqnarray}
We find that our exact solution in the ultra thin wall approximation, $\rho_B ^{\rm E}$, differs from the approximate solution, $\rho _B ^{\rm A}$ by a factor
\begin{equation}
\left| \frac{\rho _B ^{\rm E} - \rho _B ^{\rm A}}{\rho _B ^{\rm E} + \rho _B ^{\rm A}} \right| \approx 0.58 
\end{equation}
which is consistent with the size of corrections found in \cite{triscalar}. The first order correction to the baryon asymmetry arising from deviations from the ultra-thin wall regime are typically moderate to small. 
\begin{center}
\begin{table}[!h]
\begin{tabular}{|c||c|c|c|c|c|}
\hline
$L_W$ & $250/T$ & $100/T $ & $25/T$ & $10/T$ & $2.5/T$    \\
\hline
$\delta \rho _B ^{\rm E}/\rho _B ^{\rm E}$ & $0.18$ & $0.15$ & $0.14 $ & $0.14 $ & $0.12$   \\ \hline 
\end{tabular}\caption{Table of thick wall corrections to the baryon asymmetry.}\label{TWcorrections}
\end{table}
\end{center}
In figure \ref{corrections} we show the correction to the density $n_L$ for several values for the bubble wall thickness. Even though the baryon asymmetry is weakly dependant on $L_W$ in the ultra thin wall regime, for the very large values of $L_W$ we consider this dependency becomes important. The typical correction for our set of parameters is typically small but non-trivial and grows monotonically with $L_W$ as expected.  Indeed recent work has shown that bubble wall width for the electroweak phase transition in the NMSSM can have a large range of values \cite{NMSSM}. As a check on our calculations we find that the correction indeed goes to zero as $L_w \mapsto 0$ although the absolute minimum is numerically very difficult to take as it involves evaluating numerical integrals of very sharply peaked functions.
 Finally for the sake of safety we would like a direct comparison to a numerical calculation. We use the parametrized ansatz approach to solve equations (23) in Ref. \cite{winslow} in order to produce their Figure 3. This is shown in Fig \ref{winslowcheck} we are able to reproduce almost an almost identical plot (we remove the sum of densities to reduce clutter as little new information is given by this function). The small error can probably be attributed to our use of an ultra-thin wall approximation. 
\begin{figure}[htbp]
	\begin{center}
		\includegraphics[width=0.46\textwidth]{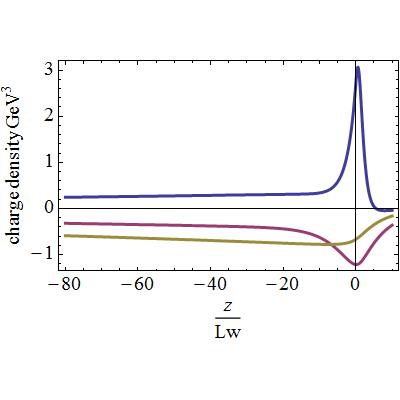} 	
		%depending on the latex compiler, you can omit the file extension
		\caption{Reproducing Winslow and Tulin's purely numerical calculation. Here the blue, magenta and golden lines correspond to the densities $n_{Q_3}$, $n_{u_3}$ and $n_{H}$ respectively}
		\label{winslowcheck}
	\end{center}
\end{figure}
\section{conclusion}\label{conclusion}
In this work we have sketched out a method to derive exact solutions to quantum transport equations in a variety of cases including large sets of QTEs and situations where there are multiple CP violating sources with non-trivially different space-time dependency. We have also produced a very quick and powerful method using a parametrized ansatz to derive the same solution. This method was found to turn cases which are cumbersome to the point of being almost intractable by standard methods to an elementary problem. Furthermore a general structure of the solution was made manifest when solving for large systems of equations. The third result of this paper was to sketch out a way of going beyond the ultra thin wall approximation. Our numerical analysis demonstrated that the correction to the baryon asymmetry coming from the smallest correction to the ultra thin wall regime can be small to moderate but nonetheless non-trivial even for moderately thick walls. It would interesting to see if other models have larger corrections from thick bubble walls.  Our numerical results analysing corrections to the fast rate approximation were consistent with those found in \cite{triscalar} which found that these corrections can be very large. Our method ensures that these corrections are taken into account while still keeping to a largely analytical framework.
\begin{acknowledgements}I would like to acknowledge Csaba Balazs, Peter Winslow and Michael Ramsey Musolf for some useful discussions relating to this work. This work was supported through the J L Williams Scholarship, APA Scholarship. Also part of this work was made possible through the Keith Murdoch Scholarship via the American Australian Association as well as the Center of Excellence for Particle Physics.
\end{acknowledgements}

\begin{appendix}
\section{Extra details in deriving the solution}
In this appendix we give extra details for how to solve the equation
\begin{equation}
a_{Q1}^i \partial ^i Q+a_{T1}^i \partial ^i T= 0 \ . 
\end{equation}
We can write $T$ as a function of $Q$ using the usual methods to write
\begin{equation}
T =\frac{1}{a_{T1}^2 }\sum _\pm \frac{1}{\kappa _\mp -\kappa _\pm} e^{\kappa _{\pm } z} \left[ \int ^z e^{- \kappa _\pm y} \left( a^i_{Q 1} \frac{\partial ^i Q}{\partial y^i}   \right) dy -\beta _i\right] \ .
\end{equation}
The first changes of variables we use is of course
\begin{equation}
h_{\pm} ^\prime = e^{-\kappa _ \pm } Q
\end{equation}
which leads to the following identities
\begin{eqnarray}
Q&=& e^{\kappa _\pm z} h^\prime _\pm \nonumber \\
Q^\prime &=& \left( \kappa _\pm h_\pm ^\prime +h^{\prime \prime }_ \pm \right) e^{\kappa _\pm z} \nonumber \\
Q^ {\prime \prime } &=& \left( \kappa _\pm ^2 h _{\pm } ^\prime + 2 \kappa _\pm  h^{\prime \prime } _\pm  + h^{\prime \prime \prime } _\pm \right) e^{\kappa _\pm z} \ .
\end{eqnarray}
This allows us to write the integrand for $T$ in terms of $h _\pm$ and remove the integral
\begin{eqnarray}
T &=& \frac{1}{a_{T1} ^2 } \sum _\pm \frac{e^{\kappa _\pm z}}{\kappa _\mp - \kappa _\pm} \int ^z e^{-\kappa _\pm y } \nonumber \\ 
&& \times \left[\left( a^0 _{Q1}+\kappa _\pm a^1 _{Q1} +\kappa ^2 _\pm a^2 _{Q1} \right) h_\pm ^\prime \right. \nonumber \\ && \left. +\left( a^1_{Q1} + 2 a^2 _{Q1} \kappa \pm \right) h _\pm ^{\prime \prime } + a^2 _{Q1} h_\pm ^{\prime \prime \prime } \right] \nonumber \\ &=& \frac{1}{a_{T1} ^2 } \sum _\pm \frac{e^{\kappa _\pm z}}{\kappa _\mp - \kappa _\pm} \nonumber \\
&& \times \left[\left( a^0 _{Q1}+\kappa _\pm a^1 _{Q1} +\kappa ^2 _\pm a^2 _{Q1} \right) h_\pm  \right. \nonumber \\ && \left. +\left( a^1_{Q1} + 2 a^2 _{Q1} \kappa \pm \right) h _\pm ^{\prime  } + a^2 _{Q1} h_\pm ^{\prime \prime } \right] \ .
\end{eqnarray}
To remove the exponents we use the change of variables
\begin{equation}
j _\pm = e^{\kappa _\pm z} h _\pm
\end{equation} 
from which we can derive the usual identities
\begin{eqnarray}
h_\pm &=& e^{-\kappa _\pm z} j _\pm \nonumber \\
h _\pm ^\prime &=& \left( j_\pm ^\prime - \kappa _\pm j _\pm \right) e^{-\kappa _\pm z} \nonumber \\
h_ \pm ^{\prime \prime } &=& \left( j^{\prime \prime } _\pm - 2 \kappa _\pm j _\pm ^\prime +\kappa _\pm ^2 j _\pm \right) e^{- \kappa _\pm z} \ .
\end{eqnarray}
We can also use the second line in the above equation to derive an identity that relates $Q$ to both new variables $j_\pm$ and its derivative
\begin{equation}
Q = j _\pm ^\prime - \kappa _\pm j _\pm \ .
\end{equation}
Our expression for $T$ is now free of any exponentials
\begin{equation}
T = \frac{1}{a_{T1}^2} \sum _\pm \frac{1}{\kappa _\mp - \kappa _\pm } \left[ a^0 _{Q1} j _\pm + a_{Q1} ^1 j _\pm ^\prime + a^2 _{Q1} j _\pm ^{\prime \prime } \right] \ .
\end{equation}
We would like to write both $T$ and $Q$ in terms of a single variable. This task is achieved by the choice
\begin{equation}
k = e^{\kappa _\mp z} \int ^z e^{-\kappa _\mp y} j _\pm dy
\end{equation}
which can be inverted to write
\begin{equation}
j_\pm = k^\prime - \kappa _\mp k \ .
\end{equation}
Both expressions for $Q$ are now satisfied
\begin{eqnarray}
Q&=& j_+^\prime-\kappa _+ j_+ \nonumber \\ 
&=& \left( k^{\prime \prime } - \kappa _- k ^\prime \right) - \kappa _+ \left( k^\prime - \kappa _- k \right) \nonumber \\
&=& k^{\prime \prime } - (\kappa _+ + \kappa _- ) k^\prime +\kappa _+ \kappa _- k
\end{eqnarray}
and 
\begin{eqnarray}
Q&=& j_-^\prime-\kappa _+ j_- \nonumber \\ 
&=&  \left( k^{\prime \prime } - \kappa _+ k ^\prime \right) - \kappa _- \left( k^\prime - \kappa _+ k \right) \nonumber \\
&=& k^{\prime \prime } - (\kappa _+ + \kappa _- ) k^\prime +\kappa _+ \kappa _- k \ .
\end{eqnarray}
We would like to write $\kappa _\pm$ in terms of the coefficients $a_{X1} ^i$ 
\begin{equation}
\kappa _\pm = \frac{-a_{T1} ^1 \pm \sqrt{(a_{T1}^1)^2}-4a^0 _{T1}a^2_{T1}}{2a_{T1}^2}
\end{equation}
which leads to the identities
\begin{eqnarray}
\kappa _+ +\kappa _- &=& -\frac{a^1 _{T1}}{a^2 _{T1}} \nonumber \\ 
\kappa _+\kappa _- &=& \frac{a^0 _{T1}}{a^2 _{T1}} \ .
\end{eqnarray}
Rescaling $k$ we get
\begin{equation}
Q= a^2 _{T1} k^{\prime \prime} + a^1 _{T1} k^\prime + a^0 _{T1} k
\end{equation}
as before. Finally substituting in our rescaled expression for $k$ we get
\begin{eqnarray}
T&=& \sum _\pm \frac{1}{\kappa _\mp - \kappa _\pm}\left[ a^2 _{Q1} k^{\prime \prime \prime} + \left( a^1 _{Q1} - \kappa _\mp a^2 _{Q1} \right) k^{\prime \prime} \right. \nonumber \\ && \left. \left( a^0 _{Q1} -\kappa _ \mp a^1 _{Q1} \right) k^\prime - a^0 _{Q1} \kappa _\mp k \right] \nonumber \\ 
&=& -a^2 _{Q1} k^{\prime \prime } - a^1 _{Q1} k^\prime - a^0 _{Q1} k 
\end{eqnarray}
as required.
\section{a note on extra integration constants}\label{note}
Throughout this work we have ignored extra integration constants that are produced in the process of reducing a set of QTEs down to a single differential equations. The contexts in which we have ignored these constants vary but throughout we reference this appendix since the reason is the same.  To demonstrate this let us consider the simplest non-trivial example. Let us have a system which can be described by the following set of QTEs
\begin{eqnarray}
a_{11}^i \partial ^i X_1 + a_{21} ^i \partial ^i X_2 &=& 0 \nonumber \\
a_{12} ^i \partial ^i X_1 + a_{22} ^i \partial ^i X_2 &=& \Delta (z) \ .
\end{eqnarray}
Solving the first equation for $X_1$ after using the usual tricks we have
\begin{eqnarray}
X_1 &=& -a_{21}^i \partial ^i k +\frac{1}{a_{11}^2}\sum _\pm \frac{e^{\kappa _\pm z}}{\kappa _\mp - \kappa _\pm}  e^{-\kappa _\pm z} \beta _\pm \nonumber \\
 & \mapsto &  -a_{21}^i \partial ^i k +\sum _\pm e^{\kappa _\pm z} \beta _\pm \nonumber \\
 X_2 &=& a_{11} ^i \partial ^i k \ .
\end{eqnarray}
The extra integration coefficients $\beta _\pm $ should immediately rouse suspicion since the system we started with was a set of two coupled second order differential equations and such a system is completely specified by four boundary conditions for every region it is solved in. Therefore a solving the equations including the use of variable changes should not introduce the need for more initial conditions. As an analogy we could make a strange change of variables that shifts $Q$ by a constant $c$. If we solve the system we should find that consistency demands that $c=0$ (rather than another condition being produced that specifies $c$) or that $c$ is redundant. These are the same requirements that will be imposed on $\beta _\pm$. Before we demonstrate the $\beta _\pm =0$ let us first show that they can be removed by another variable change. Specifically one can take
\begin{equation}
k=k_2+f
\end{equation}
where $f$ is any function that satisfies the equations
\begin{eqnarray}
-a_{21} ^i \partial _i f &=& -\sum _\pm e^{\kappa _\pm z} \beta _\pm \nonumber \\ 
a_{11}^i \partial ^i f &=&0 \ .
\end{eqnarray}
The solution is then
\begin{equation}
f=\sum _\pm A_\pm e^{\kappa _\pm }
\end{equation}
with
\begin{equation}
A_\pm = \frac{\beta _\pm}{a_{21} ^i \kappa _\pm ^i} \ . 
\end{equation}
One can then just substitute the the functions $Q$ and $T$ in terms of $k_2$ into the second equation and derive the usual solution. For completeness we nonetheless consider the case where we keep $\beta _\pm $ in the set of equations.  Substituting these solutions into the second equation gives
\begin{eqnarray}
&& (a^i _{22} a^j _{11} -a_{12}^ia_{21}^j )\partial ^{i+j}k +   \sum _\pm (a_{12} ^i \kappa ^i _\pm )e^{\kappa _\pm z}  \beta _\pm = \Delta (z) \nonumber \\
&\mapsto &(a^i _{22} a^j _{11} -a_{12}^ia_{21}^j )\partial ^{i+j}k   + \sum _\pm B _\pm e^{\kappa _\pm z} \beta _\pm = \Delta (z) \nonumber \\ \label{extraeq2}
\end{eqnarray}
Treating the term proportional to $\exp{\kappa_\pm z}$ as a inhomogeniety one could naively write 
\begin{equation}
k=\sum _{i=0}^4 x_i e^{\alpha _iz} \left( \int ^z dye^{-\alpha _i y}\left[ \Delta (y) -B_\pm \beta _\pm e^{\kappa _\pm y}\right]-\beta_i \right) \ .
\end{equation}
We can immediately see that $\kappa _{+(-)}=0$ in the broken (symmetric) phase in order to have a well behaved function at infinity. As for the $\beta _-$ term let us try and derive the coefficients $x_i$ and we will find an inconsistency unless $\beta _-=0$ (we expect such an inconsistency due to the arguments given earlier). Demanding that the coefficients of $\Delta \prime (z),\Delta ^{\prime \prime }(z)$ and $\Delta ^{\prime \prime }$ are null while the coefficient of $\Delta (z)$ is unity. This gives
\begin{eqnarray}
\sum _i x_i &=&0 \nonumber \\
\sum _i \alpha _i x_i &=& 0 \nonumber \\
\sum _i \alpha _i^2 x_i &=& 0 \nonumber \\
\sum _i \alpha _i ^3 x_i &=& (a_{11}^2a_{22}^2-a_{12}^2a_{21}^2)^{-1} \ , \label{condsdelta}
\end{eqnarray} 
as in the usual case. However,  we now have the additional condition that the coefficient of $\exp{\kappa _-z}=\beta _-B_-$ which gives the extra condition
\begin{equation}
\sum _i \frac{x_i}{\kappa _- -\alpha _i} (a^k _{22}a^j _{11}-a^j_{12}a^k_{21})\kappa _- ^{i+j}=1
\end{equation}
which gives an overdetermined set of equations unless $\beta _- =0$. Alternatively we could make use of the last line in equation (\ref{condsdelta}) to write 
\begin{eqnarray}
k&=&\sum _{i=0}^4 x_i e^{\alpha _iz} \left( \int ^z dye^{-\alpha _i y}\left[ \Delta (y) -A_iB_\pm \beta _\pm e^{\kappa _\pm y}\right]-\beta_i \right) \nonumber \\
\end{eqnarray}
with
\begin{equation}
A_i = \alpha _i (\kappa _- - \alpha _i )a_{12}^i a_{21}^j \frac{\kappa _{i+j}}{a_{12}^2 a_{21} ^2 - a_{11}^2 a_{22}^2} \ . 
\end{equation}
Upon substituting this solution back into Eq. (\ref{extraeq2}) we indeed get the equation being satisfied.  However, $X_1$ and $X_2$ are now independent of $e^{\kappa _-z}$. So depending on your treatment of $\beta _-$, it is either zero for self-consistency or irrelevant in that it is a redundancy in the theory with now physical consequences. For our purposes it is most convenient to set it to zero.
\end{appendix}


\begin{thebibliography}{99}
\bibitem{cosmicphase}
Andrei D. Linde Rep. on Prog. in Phys. {\bf 42} .3 (1979)
\bibitem{multitude1}
M. Trodden (1998).  Rev. of Mod. Phys. {\bf 71} (5) [arXiv:hep-ph/9803479]
\bibitem{multitude2}
R. Allahverdi, and A. Mazumdar.  New Journ. of Phys. {\bf 14} 12 (2012)
\bibitem{multitude3}
Satoru  Inoue, Grigory  Ovanesyan, and  Michael  J.  Ramsey-Musolf [arXiv:1508.05404v]
\bibitem{multitude4}
T. Liu, M. J. Ramsey-Musolf, and J. Shu. Phys. rev. lett. {\bf 108} 22 (2012) [arXiv:hep-ph/1109.4145]
\bibitem{electroweakbaryogenesis}
Cohen, Andrew G., D. B. Kaplan, and A. E. Nelson. Annual Review of Nuclear and Particle Science {\bf 43} 1 (1993)
\bibitem{wmap}
S. Eidelman
et al. [Particle Data Group Collaboration], Phys. Lett. B
{\bf 592}, 1 (2004).
[3] D. N. Spergel et  al. [WMAP Collaboration],  Astrophys. J. Suppl.
{\bf 148},  175 (2003)[arXiv:astro-ph/0302209]
\bibitem{phasetransition}
M. Quiros,  (1999), arXiv:hep-ph/9901312
\bibitem{CTPformalism1}
 J. Schwinger, J. Math. Phys.
{\bf 2}, 407 (1961);
\bibitem{CTPformalism2}
K. T. Mahanthappa, Phys. Rev.
{\bf 126}, 329 (1962);
\bibitem{CTPformalism3}
P. M. Bakshi and K. T. Mahanthappa, J. Math. Phys.
{\bf 4}, 1 (1963);
\bibitem{CTPformalism4}
L. V. Keldysh, Zh. Eksp. Teor. Fiz.
{\bf 47}, 1515 (1964) 
\bibitem{CTPformalism5}
R. A. Craig, J. Math. Phys.
{\bf 9}, 605 (1968);
\bibitem{CTPformalism6}
K. c. Chou, Z. b. Su, B. l. Hao and L. Yu, Phys. Rept.
{\bf 118}, 1 (1985).
\bibitem{CPVsource}
A. Riotto, Phys. Rev. D
{\bf 58}, 095009 (1998) [arXiv:hep-ph/9803357]
\bibitem{relaxation}
Christopher Lee, Vincenzo Cirigliano, and Michael J. Ramsey-Musolf. Phys Rev D {\bf 71} 7 (2005)
\bibitem{triscalar}
 Vincenzo Cirigliano et al. Phys. Rev. D {\bf 73} 11 (2006)
\bibitem{supergauge}
Daniel JH Chung, JHEP {\bf 2009} 12 (2009)
\bibitem{diffusion}
 M. Joyce, T. Prokopec and N. Turok, Phys. Rev. D
{\bf 53}, 2930 (1996) [arXiv:hep-ph/9410281].
\bibitem{gammasscorrections}
 P. Huet and A. E. Nelson, Phys. Rev. D
{\bf 53}, 4578 (1996) [arXiv:hep-ph/9506477]
\bibitem{weaksphaleronseed1}
 M. Carena, M. Quiros, M. Seco and C. E. M. Wagner, Nucl. Ph
ys. B
{\bf 650}, 24 (2003)[arXiv:hep-ph/0208043].
\bibitem{weaksphaleronseed2}
J. M. Cline, M. Joyce and K. Kainulainen, JHEP
{\bf 0007}, 018 (2000) [arXiv:hep-ph/0006119]
\bibitem{bubblewall}
J. M. Moreno, M. Quiros and M. Seco, Nucl. Phys. B
{\bf 526}, 489 (1998) [arXiv:hep-ph/9801272]
\bibitem{kfactor}
 H. A. Kramers, Physica,
{\bf 7}, 284 (1940)
\bibitem{weaksphaleron1}
D. Bodeker,  G. D. Moore and K. Rummukainen,  Phys. Rev. D
{\bf 61},  056003 (2000)
[arXiv:hep-ph/9907545]; 
\bibitem{weaksphaleron2}
G. D. Moore and K. Rummukainen, Ph
ys. Rev. D {\bf 61}, 105008 (2000)  [arXiv:hep-ph/9906259]  ;  
\bibitem{weaksphaleron3} 
G.  D.  Moore,  Phys.  Rev.  D {\bf 62},  085011  (2000)[arXiv:hep-ph/0001216]
\bibitem{strongsphaleron}
Moore, Guy D. "Sphaleron rate in the symmetric electroweak phase." Physical Review D {\bf 62} 8 (2000)
\bibitem{NMSSM}
J. Kozaczuk, S. Profumo, L. Haskins, and C. L. Wainwright, JHEP {\bf 1} (2015)[arXiv:1407.4134]
\bibitem{winslow}
P. Winslow and S. Tulin [arXiv:1105.2848]
\end{thebibliography}
\end{document}